\documentclass[12pt]{iopart}
\usepackage[square, comma, numbers, sort&compress]{natbib}
\usepackage{graphicx}
\usepackage{color,xcolor}
\usepackage{bm}
\usepackage{float}
\usepackage[normalem]{ulem}

\newcommand{\blue}{\color{black}}

\begin{document}

\title{Effects of transverse geometry on the thermal conductivity of Si and Ge nanowires}

\author{Hadi Rezaie Heris, Movaffaq Kateb, Sigurdur I.\ Erlingsson and Andrei Manolescu}
\address{Department of Engineering, School of Technology, Reykjavik University, Menntavegur 1, IS-102 Reykjavik, Iceland}



\begin{abstract}
We explore the effects of geometry on the thermal conductivity ($\kappa$) of  silicon and germanium nanowires, with lengths between 10-120 nm and diameters up to 5-6 nm.  To this end we perform molecular dynamics simulations with the LAMMPS software, using Tersoff interatomic potentials. We consider nanowires with polygonal cross section and we discuss the effect of the transverse geometry on the thermal conductivity.  We also consider tubular (hollow) nanowires and core/shell combinations of Si/Ge and Ge/Si, and we compare the heat transport of the core/shell structure with that of the separated core and shell components.


\end{abstract}



\section{Introduction}

Semiconductor nanowires have a great potential in several fields of modern technology, such as nanoelectronics \citep{duan2001indium,wang2006piezoelectric,zhao2004quantum,xu2010self}, lasers \citep{duan2003single}, solar cell \citep{tian2007coaxial}, or thermoelectrics \citep{boukai2008silicon}. For many of these applications understanding the thermal conductivity of nanowires is essential. Nanoelectronic devices demand a
high thermal conductivity and heat sinks to dissipate heat from nano chips \citep{schelling2005managing}, while a low thermal conductivity is important for achieving a high thermoelectric figure of merit \citep{zhao2014ultralow,snyder2011complex}.  

Due to the small size and large surface to volume ratio the thermal conductivity of nanowires is substantially different from that of the bulk material. 
That can be explained by the limited phonon scattering due to finite size confinement, by the increased scattering imposed by the boundary conditions, or by the quantization of phonon transport.  For example, in nanowires of diameter smaller than 20~nm, the phonon dispersion is modified due to phonon confinement, causing a significant decrease of the phonon group velocities \citep{balandin1998significant,khitun1999modification}. Experimentally, it has been reported that the thermal conductivity of individual single crystalline Si nanowires is more than two orders of magnitude lower than the bulk value \citep{li2003thermal}. The geometry plays an important  role in determining the surface to volume ratio of nanowires. Several computational \cite{nika2013thermal,cocemasov2015phonon,hu2011significant} and experimental \cite{brinzari2017ultra} studies show a relation between heat transport properties of nanowires and specific cross-sectional shapes. 

Barriers or geometric shapes can be engineered in nanostructured materials to modify the scattering of phonons.  In particular core/shell nanowires have attracted great interest for several properties, including electron and phonon scattering, and  heat and charge current through different channels \citep{lauhon2002epitaxial,Hu2010,torres2018conductance,heris2020thermoelectric}. Core/shell nanowires provide more degrees of freedom for tuning their properties compared to the  uniform nanowires. Core/shell nanowires can also have high carrier mobility \citep{xiang2006ge} and very low phonon thermal conductance due to the enhancement of surface scattering \citep{hu2011significant}, which are essential attributes for reaching higher thermoelectric efficiency. Previous studies addressed the effects of surface construction and geometry  on nanowires's intrinsic electrical conductivity \citep{ng2008geometry}, thermal conductivity \citep{tessema2017effect}, mechanical \citep{ji2007coupled} and thermoelectric properties \citep{liang2010geometry}. The growth of semiconductor nanowires with specific cross-section is  a current research focus in fabrication of nanowires, since many of their properties depend on the geometry \citep{yang2005chemistry,zou2007growth}, such as hexagonal \citep{ren2004hexagonal,noborisaka2005catalyst,yuan2015antimony}, square \citep{qin2008template,he2019hexagonal} and triangular \citep{pennelli2006fabrication,yuan2015antimony,zou2007growth} cross-section.

Great effort have been made to study heat transport through core/shell nanowires. These studies show that  a deposition of a thin shell \cite{prasher2006thermal}, the interface roughness \cite{chen2012impacts} and depression and localization of long-wavelength phonon modes reduce the thermal conductivity of  nanowires \cite{chen2011phonon}. Also the drastic reduction of thermal conductivity can be achieved through cross-section modulation \cite{hu2011significant,nika2013thermal}. The finite size effects related to the length of the nanowire have been obtained for silicon nanowires with square cross-section and lengths between 6--54~nm \cite{wang2009thermal} and for Si/Ge core/shell nanowire with the same square geometry \cite{Hu2010}.

Among the variety of theoretical approaches molecular dynamics (MD) has received considerable attention for the study of phonon behaviour. This is mainly due to the fact time-scale and spatial resolution in MD simulation perfectly fits for observation of atomic vibrations. Since the pioneering work of Green and Kubo \citep{green1954,kubo1957}, several algorithms for calculating $\kappa$ within the MD framework  have been developed.  In the Green-Kubo approach ensemble average of the instantaneous heat flux auto-correlation is related to $\kappa$ through fluctuation-dissipation theorem. Note that Green-Kubo algorithm does not require a temperature gradient and thus it is referred to as equilibrium molecular dynamics (EMD) \cite{Volz99}. The method has been already applied to nanowires of various polygonal cross-sections and infinite length \citep{ma2016nano,soleimani2018comparative,ponomareva2007thermal,dong2018equivalence}.
 The lack of temperature gradient initiated a debate whether EMD method is proper for one-dimensional solids or should be modified (cf.\ Ref.~\citep{khadem2013} and refs.\ therein). However, the major problem with EMD was its computation cost, namely for the auto-correlation function to be convergent. This motivated the non-equilibrium molecular dynamics (NEMD) approach based on imposing a temperature gradient and utilizing Fourier's law to calculate $\kappa$. Since then several studies have been devoted to understanding the origin of difference in $\kappa$ determined by EMD and NEMD approaches (cf.~Refs.\citep{schelling2002,khadem2013,li2019}). The temperature gradient in NEMD is achieved by thermostating hot and a cold segments at different temperatures or adding and removing an equal amount of heat to these segments. Thus, it is generally believed that NEMD approach mimics better the experimental condition. However, the NEMD approach often presents a non-linear temperature profile in the neighborhood of the hot and cold segments. Whether or not the non-linear portions must be included in determining $\kappa$ is still an ongoing debate \citep{li2019}. Besides, the NEMD approach gives large fluctuations temperature/heat flux unless e.g.\ imposing a large temperature gradient to have the heat flux distinguishable from the noise \citep{Muller-Plathe97}. Then the large heat flux requires a longer simulation to be converged. M\"{u}ller-Plathe \citep{Muller-Plathe97} introduced the reverse-NEMD (rNEMD) approach in which the heat flux is imposed by exchanging kinetic energy between the hot and cold segments, and produces a temperature gradient. This way the heat flux is already known from the amount of exchanged energies. Besides, the rNEMD method conserves the energy and thus overcomes the dependency on a specific ensemble.

In the present paper we calculate with MD simulations the heat conductivity of silicon nanowires with different shapes of the cross-section, hollow nanowires, and Si/Ge core/shell nanowires of triangular shape.  We use the rNEMD method to investigate the thermal conductivity in tubular and core/shell nanowires  with different geometries of the cross-section: cylindrical, hexagonal, square, diamond, and triangular.
We begin with a theoretical background of rNEMD in Section 2. Then, Section 3  describes our model and method to calculate the heat flux and thermal conductivity. In Section 4 we discuss the thermal conductivity variation with the nanowires length and cross-sectional geometry, and also the thermal conductivity variation from solid nanowires to tubular ones with different geometries. We also compare the thermal conductivity of Si/Ge and Ge/Si core/shell nanowires with triangular cross-section. Finally, the conclusions are reported in Section 5.


\section{Theoretical background on the reverse non-equilibrium molecular dynamics (rNEMD) approach}

We use the rNEMD method by M\"{u}ller-Plathe  to compute the heat flux and thermal conductivity of our models of nanowires \cite{Muller-Plathe97}. 
In this approach the heat flux density $Q$ is defined as the amount of energy transferred in a given time through a surface of a given area which is perpendicular to the flux direction. The thermal conductivity $\kappa$ is calculated by Fourier's law:
\begin{equation}
    Q=- \kappa \nabla T
    \label{fourier law1} \ ,
\end{equation}
where $\nabla T$ is the gradient of the temperature $T$ determined from atomic velocities.  If we take $z$ direction as the direction of the temperature gradient along the nanowire, we can define the thermal conductivity as
\begin{equation}
  \kappa = - \lim_{\partial T / \partial z \to 0} \lim_{t \to \infty} \frac{\langle Q_{z}\rangle} {\langle \partial T / \partial z\rangle} \ ,
    \label{fourier law}
\end{equation}
where $t$ is the simulation time, in practice longer than the relaxation time of the atomic structure after the initial conditions. 
In order to make sure the determined $\kappa$ exactly belongs to the nanowire axis, the M\"{u}ller-Plathe algorithm requires extra considerations. 

The nanowire must be divided into $n$ segments along its axis, $n$ being an odd integer, and each segment having the same thickness, volume, and number of atoms. The temperature in the segment $S_k,\ k=1,2,\dots,n$, is given by
\begin{equation}
\frac{3}{2}N_{k}k_{\rm B}T_{k}=\frac{1}{2} \sum_{i\in S_k}^{N_{k}} m_{i} \mathbf{v}_{i}^2 \ ,
\label{temp}
\end{equation}
where the sum extends over the $N_{k}$ atoms contained in $S_k$, with masses $m_{i}$, velocity vectors $\mathbf{v}_{i}$, $k_{\rm B}$ being Boltzmann's constant. For the case of periodic boundary conditions, the hot segment is almost located in the middle of the nanowire, corresponding to $S_{(n/2)+1}$, and the cold segment is created at the first $S_{1}$. Note that due to the periodic boundary condition the heat flows from the hot segment to both ends of the wire. The temperature gradient is produced by exchanging the $\mathbf{v}_{i}$ of the most energetic atoms situated in the $S_{1}$ with the least energetic atom situated in the $S_{(n+1)/2}$. Considering how often the exchange process is performed and how many atoms are included in each exchange step, one can control $\Delta T$ between hot and cold segments. This allows producing a linear $\Delta T$ between the hot and cold segments \citep{Muller-Plathe97}, in contrast to other NEMD methods. 
The heat flux per unit area and time is calculated as 
\begin{equation}
  Q_{z}=\frac{1}{2A}\frac{dE}{dt} \ ,
    \label{fourier law2}
\end{equation}
where $A$ is cross-sectional area of the wire, counted twice because of of the heat flow in both directions, from the central hot segment to the lateral cold segments. Assuming the validity of a linear response (Fourier's law), the thermal conductivity is calculated by combining Eqs.\ (\ref{fourier law1}) and (\ref{fourier law2}).
 
To check the spatial distribution of the heat flux within the nanowire cross-section we utilized atomistic values along the nanowire axis. The per-atom heat current ($\mathbf{q}$) is given by
\begin{equation}
    \mathbf{q}_{i}=\frac{1}{\Omega}\frac{d}{dt} \mathbf{r}_iE_i,
    \label{eq:transport}
\end{equation}
where $\Omega$ is atomic volume, $E_i$ is the total (potential and kinetic) energy of particle $i$ and $\mathbf{r}_i$ is its position vector. Although there exist a temperature gradient $\Delta T$, at the atomistic level we do not need a temperature gradient because $\frac{d}{dt}E_i$ considers variation of per-atom kinetic energy through time. With $\mathbf{q}_{i}$ being the microscopic or atomistic or per-atom value, macroscopic equivalent can be obtained by $Q_{z}=\sum_{i}q_{i}^{z}$.  For a pair potential Eq.~(\ref{eq:transport}) can be expressed as
\begin{equation}
    \mathbf{q}_{i}=\frac{1}{\Omega}[E_i\mathbf{v}_i+\frac{1}{2}\sum_{j} (\mathbf{F}_{ij}\cdot \mathbf{v}_i)\mathbf{r}_{ij}],
    \label{eq:pair}
\end{equation}
where $\mathbf{v}_i$ is the velocity vector of the particle and $\mathbf{F}_{ij}$ and $\mathbf{r}_{ij}$ are force and separation vector, respectively, between particle $i$ and its neighbor(s) $j$. Note that Eq.~(\ref{eq:transport}) includes $\mathbf{r}_i$ that is meant for liquids i.e.\ when energy can be transferred by movements of atoms. The second term in the right hand side of Eq.~(\ref{eq:pair}) is the virial contribution to the heat flux and thus one may rewrite Eq.~(\ref{eq:pair}) as
\begin{equation}
    \mathbf{q}_{i}=\frac{1}{\Omega}[E_i\mathbf{v}_i+\boldsymbol{\sigma}_i\mathbf{v}_i],
    \label{eq:stress}
\end{equation}
where $\boldsymbol{\sigma}_i$ is per-atom stress tensor of atom $i$. Now one can decouple $\mathbf{q}_i$ into different directions e.g.\ assuming the $z$  being parallel to the nanowire axis and direction of interest
\begin{equation}
   \mathbf{q}_{i}^{z}=\frac{1}{\Omega}[E_i\mathbf{v}_i^z+\boldsymbol{\sigma}_i^{zz}v_z+\boldsymbol{\sigma}_i^{yz}v_i^y+\boldsymbol{\sigma}_i^{xz}v_i^x],
\end{equation}
with superscript being the directions. We refer the interested readers to Ref.~\citep{surblys2019} for discussion on the stress correction when manybody potentials such as Tersoff is used.

\section{The atomistic model}

We use silicon and germanium with diamond structure as a model  systems  in  our  simulations. An atomistic model which ignores electron transport can be realistic for silicon since electrons in  the intrinsic (undoped) Si contribute very  little to the  heat  conduction compared to the phonons.  This can be achieved by MD simulation and solving Newton's equation of motion. To this end we utilized the large scale atomic/molecular massively parallel simulator (LAMMPS) from Sandia National Laboratory \cite{Plimpton95}.

 The Si nanowires were placed within a orthogonal simulation box that fits the length of nanowires, but it is several times larger than its diameter in the transverse directions, {\blue as shown in Figures \ref{fig:geo_wire} and \ref{fig:geo_tube} of the 
Appendix.}
We apply fixed boundary conditions in transverse directions ($x$ and $y$), and periodic boundary conditions along the nanowire axis ($z$ direction). 
A question that might arise here is if our nanowires can be considered as having infinite length. As pointed out previously \citep{dong2018equivalence}, and as will be discussed in the results, $\kappa$ is dependent on the actual nanowire length included in the simulation. However, the nanowire length utilized here does not correspond to the real sample size in experiment. This means that periodic boundary condition can reduce the finite size effect, but due to the phonons it can not be completely removed. 
For this reason we considered different nanowire lengths (10--120~nm) and discuss the issue further in the results. 

The nanowire axis was aligned along the [111] lattice direction that gives (111) planes at the nanowire cross-section. 
 In order to understand the trade-off between the effect of surface area and cross-section area we compared  uniform and hollow nanowires with equal cross-sections.
The  cavity, or hole, in the hollow nanowires is built with a central empty channel inside the material and again, we compare different geometries by keeping the transverse area constant. {\blue In order to generate the core-shell nanowires we replaced core atoms
with a different materials and the entire core-shell was relaxed to reach equilibrium
interatomic distances. This gives an interface similar to that of hetero epitaxial
system.}


An important requirement for MD simulation of solid materials is the 
availability of accurate interatomic potentials \citep{rohskopf2017empirical}.  Several MD simulations studies on thermal conductivity show strong dependence of these results on the empirical interatomic potential used \citep{zhou2017nonmonotonic,fan2019minimal}. 
Such potentials have mostly been developed for technologically important materials, such as silicon and germanium. 
Among them, the Tersoff potential \citep{tersoff1988new} and its variants are most commonly used \citep{khadem2013,dong2018equivalence}.  It  is a many-body potential that allows the description of covalently bonded materials, capturing the quantum mechanical nature of the bonding, and explicitly incorporating the dependence of the bond order on local environment. We compared Tersoff potential with the popular Stillinger-Weber potential and a more recent EDIP potential as well as DFT in determining relevant thermodynamic properties such as heat capacity. The results are presented in the Appendix. {\blue We utilized Tersoff \citep{tersoff1989modeling} potential for Si--Si, Ge--Ge and Si--Ge interactions.}

We proceed with a two step relaxation process. First, the nanowire is relaxed using the isothermal–isobaric ensemble (NPT) at the desired equilibrium temperature $T_{eq}$ in order to allow the atoms to expand until nearly zero stress is achieved along the nanowire axis, within 50~ps. {\blue We utilized Nose-Hoover thermostat to achieve desired temperature during the relaxation.} In the second step, the energy exchange between the hot and cold regions is performed, in order to develop a stationary temperature profile along the nanowire axis, for another 50~ps using the microcanonical ensemble (NVE).  Note that NVE allows the average temperature over all segments to remain constant and equal to $T_{eq}$. We applied different settings for the exchange to produce the temperature differences being in the range of 10-60~K between the central and the side segments. 
We used the velocity Verlet algorithm for time integration \citep{verlet1967,kateb2012} using a time step of 0.5~fs, and $\Delta T$ and consequently $\kappa$ were determined when averaged data points were converged. We used 10$^3$ samples to calculate each of the averaged $\Delta T$ points.

\begin{figure}
    \centering
    \includegraphics[width=0.7\linewidth]{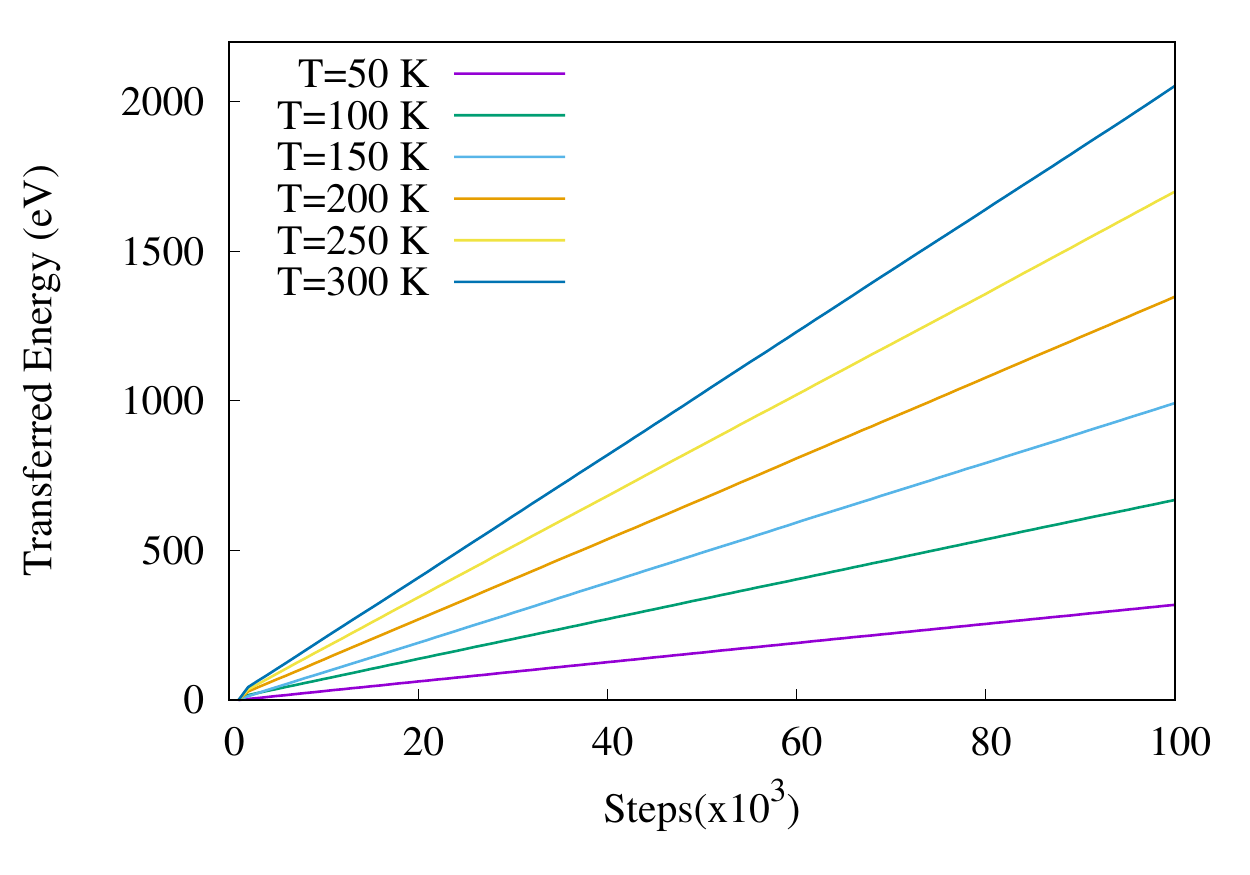}
    \caption{Cumulative energy transferred from the hot to the cold segments of the nanowire with elapsed time.  Each simulation step corresponds to 0.5 fs.}
    \label{heat-step}
\end{figure}

In Fig.\ref{heat-step} we show the cumulative energy transferred along the cylindrical nanowire axis fot temmperatures between 50 - 300 K. We can see that heat transfer increases linearly with respect to the number of time steps, and that is different for each temperature. This is an indication of the fact that steady state temperature profile is reached and thus we can use Fourier's formula for calculating the thermal conductivity. {\blue Note that below
50~K the heat flux carried by electrons is comparable with the contribution of phonons, so studies at lower temperature have to consider and both electrons
and the upper limit of the thermal conductance for phonons. This fundamental upper limit 
(a quantum of thermal conductance) 
is the maximum value of thermal conductance that an acoustic phonon mode can take \citep{angelescu1998heat,rego1998rego}. Also at higher temperatures
numerical approaches of Holland's model (extended Callaway’s model) can describe
thermal conductivity of silicon and germanium up to 1000 K easily \citep{holland1963analysis}.}
%


\section{Results and discussion}

\subsection{Length dependence of the thermal conductivity} 

 There are several reports on the dependency of thermal conductivity to the length of nanowire within MD framework \citep{lukes2007thermal,alaghemandi2009thermal}. The Fourier's law describes thermal transport in bulk materials, but it is not valid for nanowires with lengths less than phonon mean-free path. For  shorter nanowires ($L<< \lambda$), it has been shown experimentally that the thermal conductivity does not have a constant value \citep{yu2005thermal}. The thermal conductivity variation of these nanowires is strongly related to the temperature regime. At low temperature where the optical phonons are not excited and thermal current is carried ballistically by acoustic phonons, thermal conductivity can be described by Landauer formulation for phonons \citep{yamamoto2004universal,yamamoto2006nonequilibrium}. Meanwhile for temperature regime higher than 20~K thermal conductivity is not proportional to the system length and increase non-linearly with the length due to contribution of both ballistic and diffusive phonons \citep{alaghemandi2009thermal,yamamoto2004universal}.


The thermal conductivity of a pure silicon nanowire increases rapidly with nanowire length below 100~nm. This behavior is in agreement with previous MD studies \citep{wang2009thermal,hu2011thermal}. The thermal conductivity varies more slowly at higher lengths and reaches a constant value at lengths over 1000~nm. This slow variation of the thermal conductivity implies that the nanowire is in the diffusive phonon transport regime. The variation of the thermal conductivity with the length of the nanowire could be understood by considering a linear variation of the reciprocal of the thermal conductivity with respect to the reciprocal of the nanowire length, or a power law behavior of the thermal conductivity with respect to the length.

Considering the temperature regime and nanowires of different lengths, we describe our results with a power law relation. The thermal conductivity increases with respect to the length of the nanowire according to $L^\alpha$, where $\alpha$ varies between 0.30-0.48 for $L < 30$~nm, and reaches a  constant value of 0.56 for 30~nm $< L < $ 150~nm. The variation of $\kappa$ with the nanowire length can be understood as the limitation imposed to contribution of long wavelength phonons. As the nanowire length increases a wider spectrum of phonons contribute to the calculated heat flux. 
We study nanowires with different lengths along $z$ direction to see how the thermal conductivity behave when the length is comparable to the phonon mean free path. 
%
\begin{figure}
    \centering
    \includegraphics[width=0.7\linewidth]{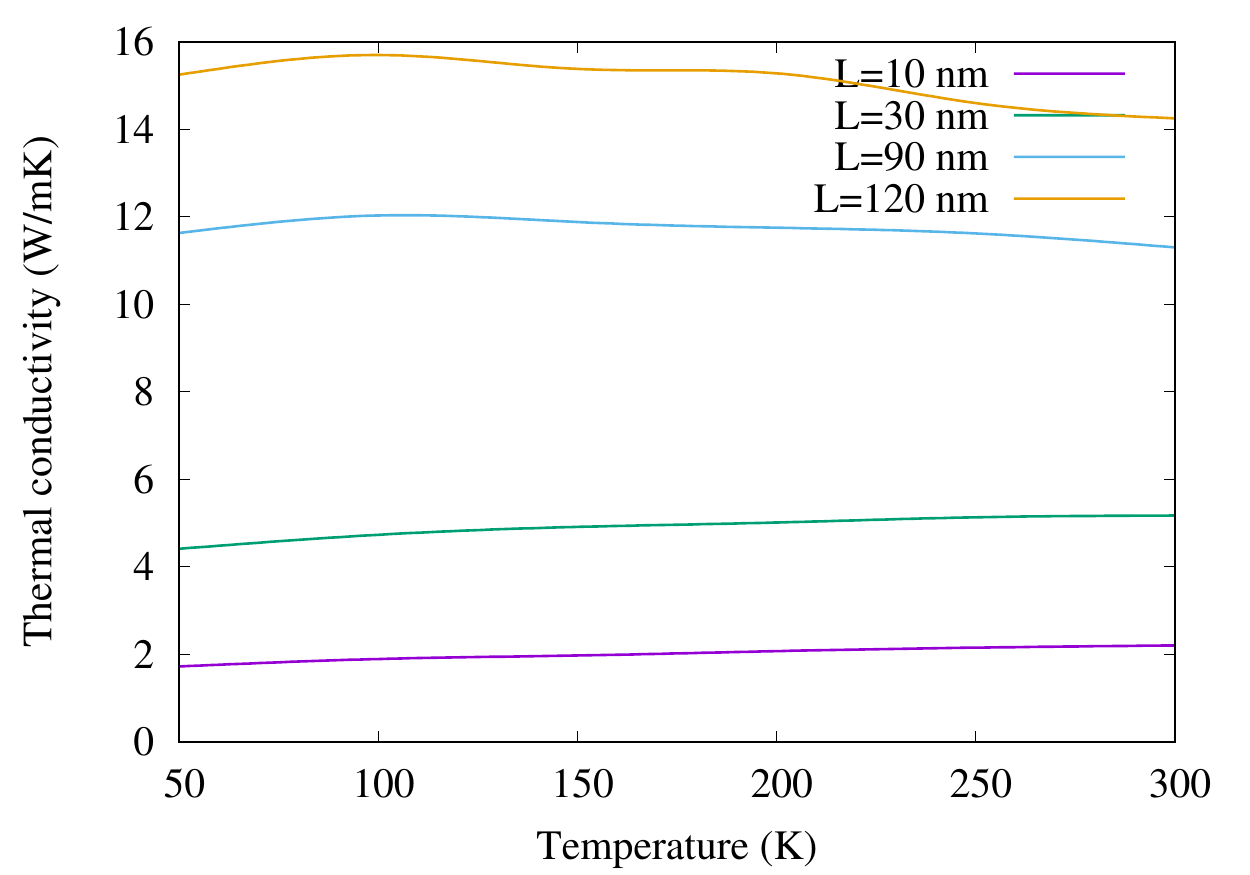}
    \caption{Variation of thermal conductivity for a cylindrical nanowire of 5.5 nm diameter with length increase}
    \label{length}
\end{figure}

Our calculated thermal conductivity is displayed in Fig.\,\ref{length}, for nanowires with length $L$ from 10 to 120~nm, having cylindrical shape. 
{\blue We can see a huge increase of the thermal conductivity by increasing the nanowire length. For instance at 100~K, the thermal conductivity $\kappa$ for the 120~nm length nanowire is almost eight times higher than for the one with 10~nm length.}

\subsection{Effect of the nanowire cross-section on the thermal conductivity} 

\begin{figure}
    \centering
    \includegraphics[width=0.7\linewidth]{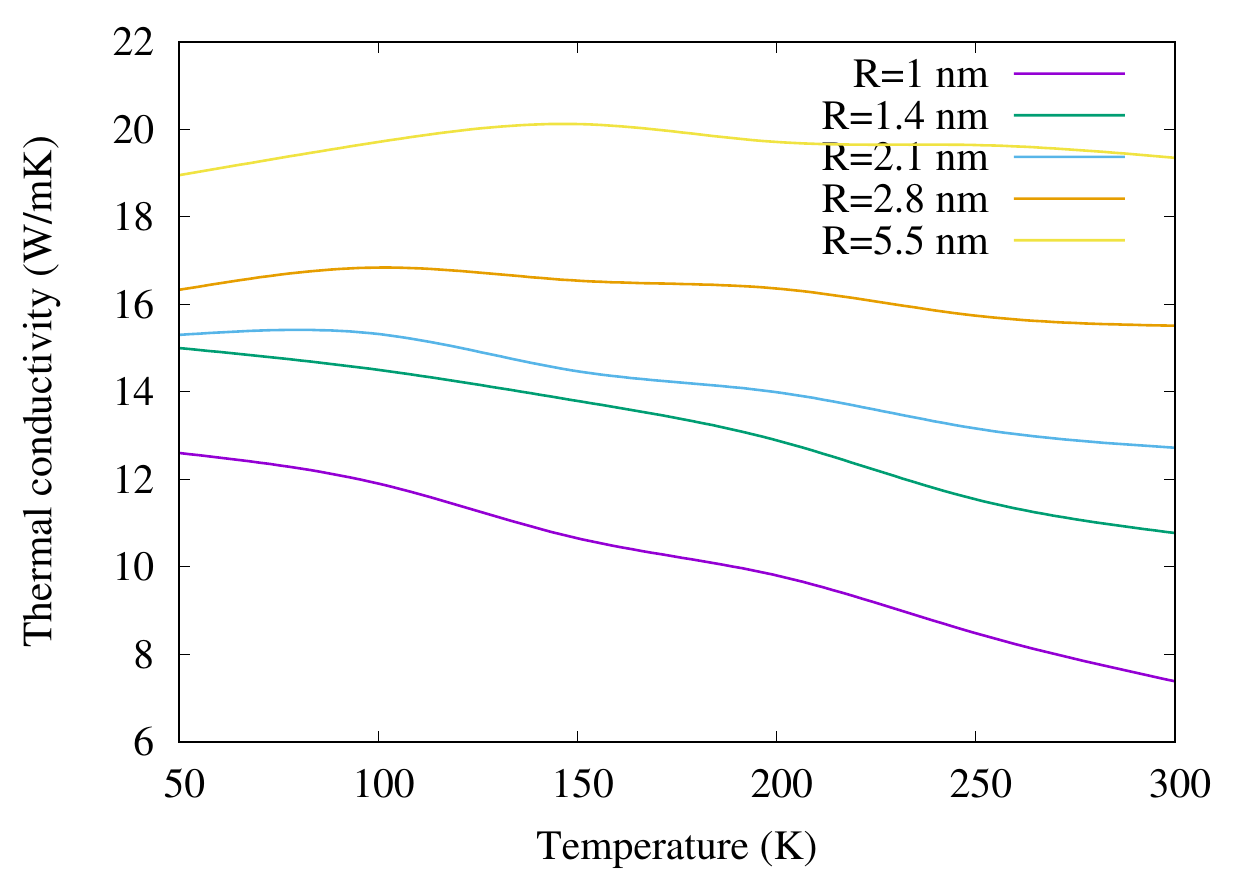}
    \caption{Thermal conductivity of cylindrical nanowire with different diameters, indicated in legend, at a constant length of 120 nm}
    \label{diameter}
\end{figure}

If we fix the length of nanowires and look at the influence of the cross-sectional area on thermal conductivity with respect to temperature, we can see an increase of the thermal conductivity with increasing the area. Although these variations are not same for each temperature, the trend is  observable. {\blue Specifically at 300~K the thermal conductivity magnitude increase more than two times by increasing radius of cylindrical nanowires from 1~nm to 5.5~nm, as shown in Fig.~\ref{diameter}.}

%

\begin{figure}[H]
    \centering
    \includegraphics[width=1\linewidth]{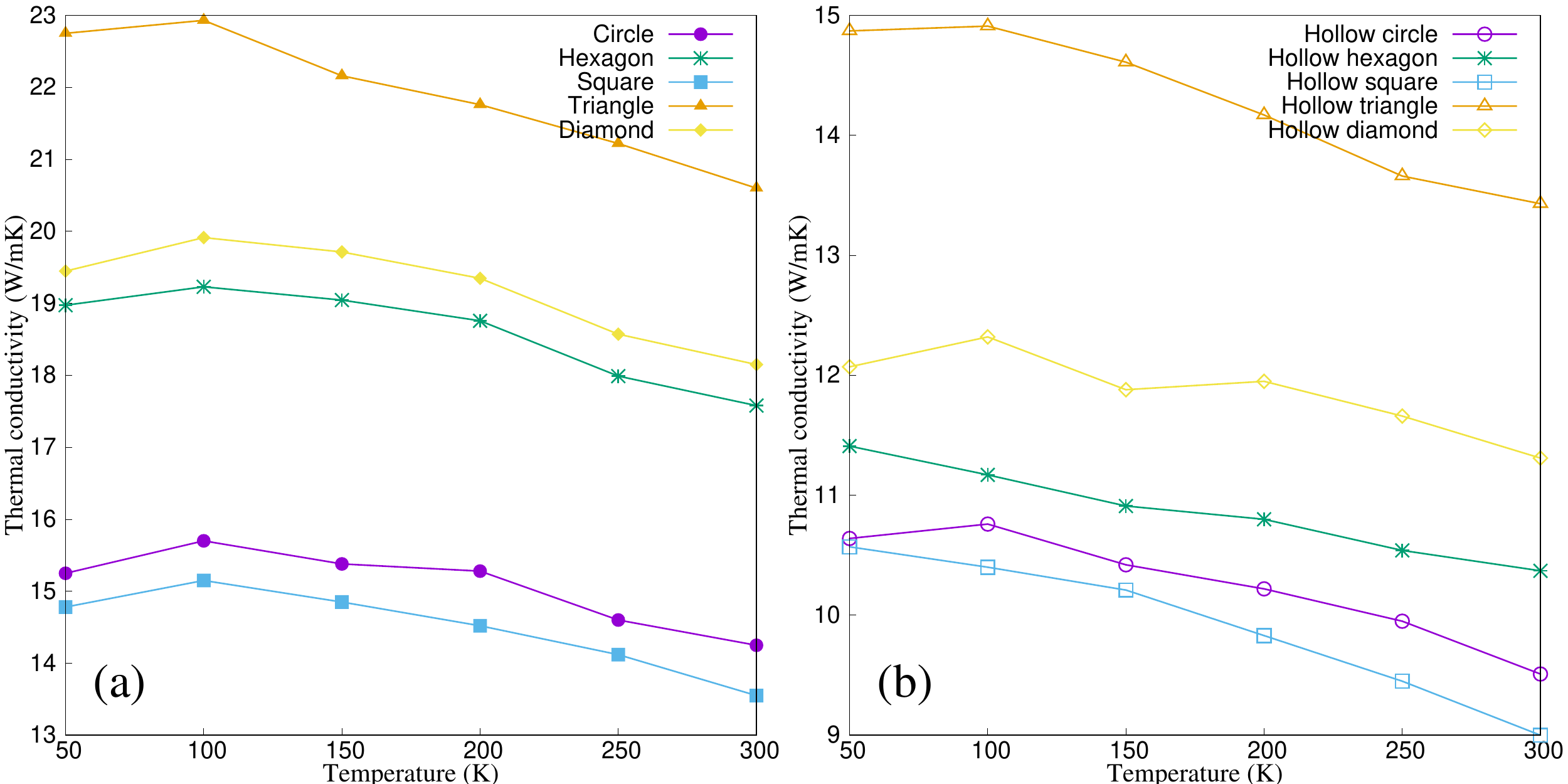}
    \caption{Thermal conductivity variation with temperature for different geometries. (a) solid nanowires with length of 120 nm and surface area of 24 nm$^2$ (b) Tubular nanowires with length of 120 nm and surface area of 24 nm$^2$ }
    \label{heat current}
\end{figure}

The dependence of the thermal conductivity and thermoelectric properties of silicon nanowires on the cross-sectional shape and surface modification had been studied before \cite{ng2008geometry,tessema2017effect,liang2010geometry}. In our present study we consider the effect of different geometries. We fix some parameters such as length, cross-sectional area, and the number of atoms in a cross section, for all shapes, and we only change shape of the nanowire cross section, from circle to triangle, diamond, hexagon and square.  The results are shown in Fig.\ \ref{heat current}(a). {\blue The thermal conductivity of silicon nanowires is about a factor of two lower than for bulk silicon \citep{li2003thermal}}. Inelastic surface scattering and finite size phonon effects contribute to the low thermal conductivity of nanowires, which is consistent with prior results \cite{lin2013thermal,liang2010geometry}. Nanowires with square and circle cross-sectional area show very close values of thermal conductivity, the values for the circular case being slightly larger for all temperatures. In some studies similar results for square and circle shaped nanowires had been reported \cite{hu2011significant,lin2013thermal,lu2006lattice}. In our study the nanowires with triangular cross-section area (with a fixed area of 24 nm$^2$) have the highest thermal conductivity of all shapes. Below that, the next values correspond to diamond and hexagon.  

{\blue The largest conductivity for triangular nanowires seems to be related to the largest flat surfaces present in this geometry.  Or, the presence of many edges, or corners of the cross section, reduce the conductivity. In other words, the thermal conductivity increases with increasing the surface-to-volume ratio, or perimeter-to-area ratio of the cross section, because our nanowires have all the same length.  Since the cross sectional area $A$ of all shapes is also the same, the smallest ratio corresponds to the cylinder ($3.5/\sqrt{A}$), followed by the hexagon ($3.7/\sqrt{A}$), square ($4/\sqrt{A}$), diamond ($4.3/\sqrt{A}$), and triangle ($4.6/\sqrt{A}$).  Our thermal conductivities obey this order, except for the the square case. Our interpretation of this exception is related to the unit cell of the material (diamond), which does not accommodate well in the square cross section of a nanowire with its length in the [111] direction. In Figure \ref{fig:geo_wire} of the Appendix one can see a zigzag distribution of the surface atoms in the square case.  The corner-to-corner distance for the square nanowire is thus smaller than for atoms with a better alignment, like in the other geometries, and the effective surface-to-volume ratio for the square geometry is actually smaller than expected. To understand better this situation, we tried to change step-wise the corner angles of the diamond shape in order to observe the evolution of the thermal conductivity to the results for the square shape. However, for many shapes in between diamond and square the relaxed structures had irregular surfaces and the results for the thermal conductivity were not stable.  We believe this is an additional indication of the geometrical mismatch between the nanowire geometry and the unit cell, at least at this small scale considered in our work.  
}

\subsection{Hollow nanowires} 

\begin{figure}[H]
    \centering
    \includegraphics[width=1\linewidth]{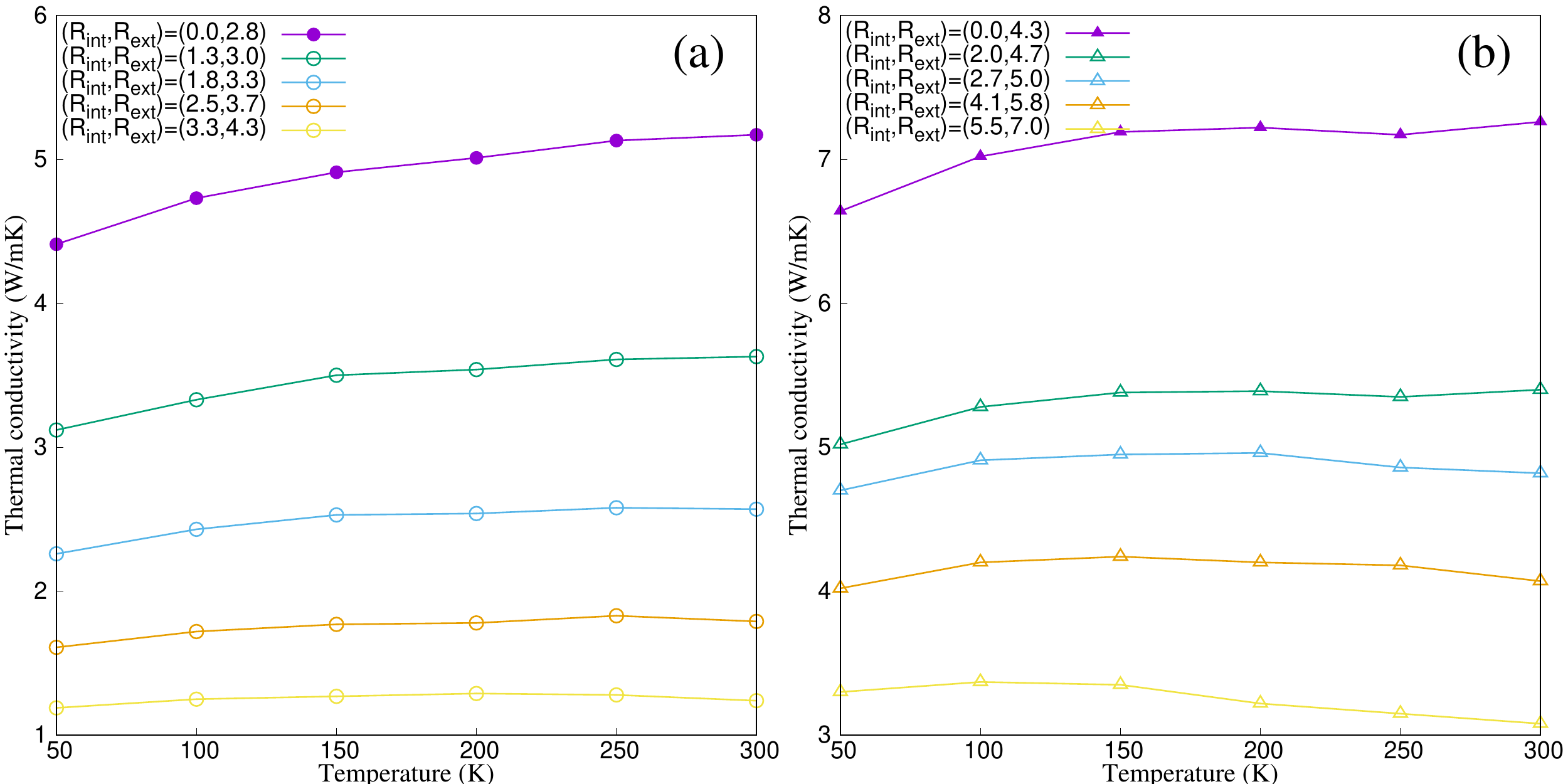}
    \caption{The thermal conductivity vs. temperature with increasing the hollow space inside the nanowires with (a) cylindrical and (b) triangular prismatic shapes. The shell area $\approx$ $24 \ {\mathrm {nm}}^2$ and the length $L=30$ nm are same for all cases.}
    
    
    \label{tubular}
\end{figure}

We are  modeling tubular nanowires by omitting atoms from the nanowire center, and see a huge decrease of the thermal conductivity values for all shapes. That is predictable due to the increase of the surface to volume ratio. 
The consequence is phonon confinement and enhanced phonon scattering, leading to lowering of the thermal conductivity in the tubular nanowires in comparison with  uniform nanowires. In the next phase by considering the same cross-sectional area, number of atoms, and nanowire length, this time we increase the empty space inside the nanowire material. Explicitly the thermal conductivity decreases with increasing the nanowire radius, as shown in Fig.\ \ref{tubular} for the circular and triangular cases. 
All the other shapes show same behavior. The larger radii (both internal and external) leads to lower thermal conductivity. The reduction for the triangular cross section can be more than 50\%, which is the largest effect for all studied shapes. 

\begin{figure}
    \centering
    \includegraphics[width=0.7\linewidth]{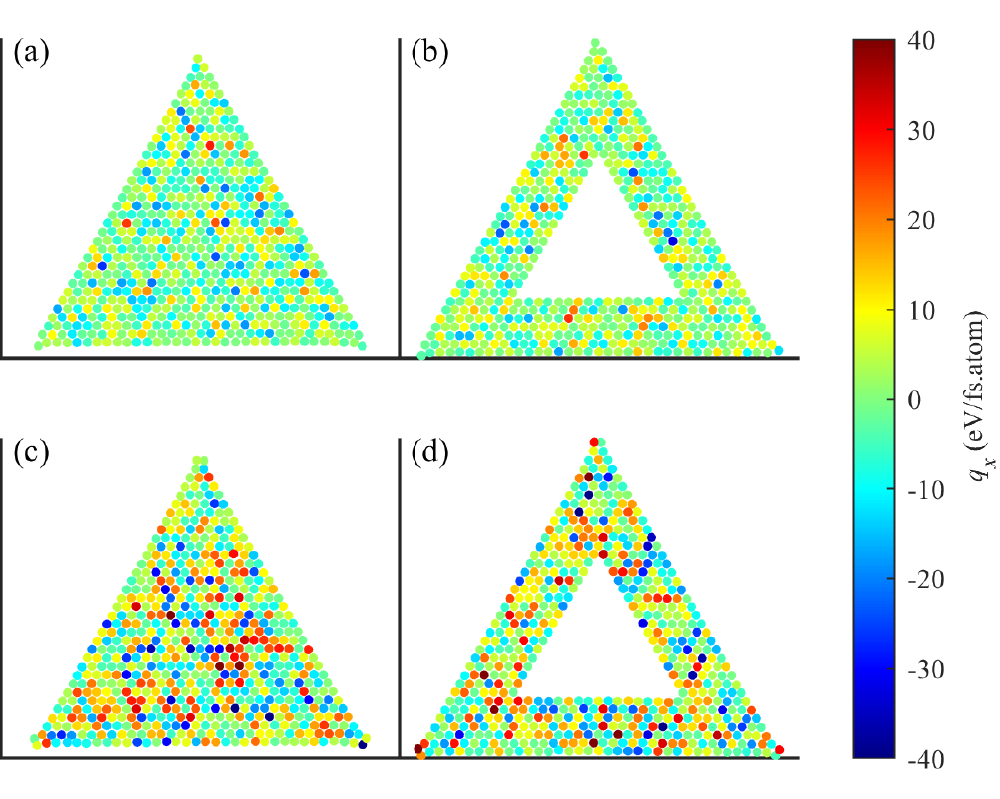}
    \caption{Distribution of longitudinal heat flux ($q_x$) within nanowire cross-section. Snapshots of (a--b) $q_x$ at a specific $t$ and $x$, (c--d) the average of $q_x$ over time and nanowire length i.e.\ $\langle q_x\rangle_{t,x}$. }
    \label{fig:flux}
\end{figure}

Next, in Fig.~\ref{fig:flux} we show a snapshot of the average of the heat flux along the nanowire axis for triangular nanowires, both with compact and hollow geometry, at temperature 100~K. The averaged heat flux was obtained by averaging over both time and half length of nanowire i.e.\ where slope of the temperature profile does not change sign. The color bar indicates the heat flux in eV/atom.fs units. 
It can be seen that there is no obvious heat channel and the distribution of heat flux fluctuate around a uniform value, within the nanowire cross section, for both geometries. Strictly speaking, however, it seems larger flux passes through the nanowire core in the absence of the hollow space. So with the empty space in the center of the nanowires we omit the main heat flux through its core, which leads to a lower thermal conductivity.

%
%

%
%

%

\subsection{Core/shell nanowires}

Core/shell nanoscale structures allow additional possibilities to engineer the thermal transport, via phonon scattering mechanisms. The control of phonon interference at the interface between the core and the shell, and the phonon scattering in different geometries, makes  core/shell nanowires a promising candidate for heat transport at nanoscale.
In principle, the thermal conductivity of independent parallel nanowires should always be larger than the conductivity of each individual nanowire (as in the case of electrical conductivity). But for a core/shell nanowire, with two different materials in contact, it is more difficult to calculate the thermal conductivity, in order to incorporate the effects of the stress along the common interface, and the different  phonon group velocities in each material. 


In order to relate the cross section shape with the variation of the  thermal conductivity, we performed three series of simulations. First the thermal conductivities for Si/Ge core/shell nanowires, between temperatures 50-300 K, were calculated for all cross sectional shapes. And then, the thermal conductivities  for  hollow  and   uniform  nanowires,  corresponding  to  the  shell  and  core separated from each other, but using the same cross sectional area, number of atoms, and length, as for the core/shell structures. The lengths of all nanowires are $30 \ \mathrm {nm}$ and the cross sectional area is $24.5 \ \mathrm {nm}^2$ for both core and shell nanowires, respectively Si/Ge core/shell total cross sectional area is  $49 \ \mathrm {nm}^2$. 

For understanding the thermal conductivity variation for each shape, we calculate the average thermal conductivity of the core ($\kappa_c$) and shell ($\kappa_s$) nanowires, $\kappa_{ave}=(\kappa_{c}+\kappa_{s})/2$, 
for each specific shape, and we compare it with the true thermal conductivity of the Si/Ge core/shell nanowire of that shape ($\kappa$). {\blue The meaning of $\kappa_{ave}$ is the thermal conductivity of the core/shell nanowire if the core and the shell would behave like independent heat transport channels (both core and shell have the same cross-sectional area). The thermal conductivities for the core, shell, and core/shell cases, and the corresponding deviation ($\kappa_{ave}-\kappa)/\kappa$ are reported in Table 1, for temperature 100 K. In all cases the true core/shell value $\kappa$ is smaller than the average value, indicating that the interface between the core and the shell reduces the heat transport.  We see that for the square and circular cross sections these differences are the smallest, less than 2\%. The largest difference is again for the triangle, and then for the diamond. }

\begin{table}[hp]
\begin{center}
\begin{tabular}{ |c|c|c|c|c| }
 \hline
 Cross section shape & Si core $\kappa_c$ & Ge Shell $\kappa_s$ & Si/Ge core/shell $\kappa$ & Deviation of $\kappa_{ave}$ \\ 
 \hline
 Circle & 4.89 & 3.46 & 4.10 & 1.8\% \\ 
 \hline
 Hexagon & 5.76 & 3.04 & 4.14 & 6.3\% \\ 
 \hline
 Diamond & 6.11 & 5.27 & 5.28 & 7.7\% \\ 
 \hline
 Square & 4.87 & 3.49 & 4.11 & 1.7\% \\ 
 \hline
 Triangle & 7.09 & 5.30 & 5.33 & 16\% \\ 
 \hline

\end{tabular}
\caption{\label{tab:table-name}Thermal conductivity of Si core, Ge shell and Si-Ge core/shell nanowires with different cross sectional shapes. The temperature is 100 K. The length of all nanowires is 30 nm.}
\end{center}
\end{table}

{\blue A significant reduction of the thermal conductivity of a Si nanowire with square cross section, if coated with a Ge thin layer, was already reported in the literature \cite{Hu2010}.  
This  effect  can be explained  by  the  strong  reduction  of  phonon  group  velocities  in 
core/shell  nanowires  due  to  the  redistribution  of  phonon  energy  spectra  and  mixing  of  high-velocity phonons from silicon with low-velocity phonons from germanium \cite{crismari2012thermal}.
Here we find out that the reduction can be larger for other geometries, especially for those with large surface-to-volume ratio, like in the triangular case.}
The high deviation from average thermal conductivity $\kappa_{ave}$ values for triangle shape can be created by phonon confinement effects imposed by the geometry. Due to this reason we chose the triangle shape for performing further simulations.

In the next phase of simulations for the Si/Ge case, the Si core cross-sectional area is $5.2 \ \mathrm {nm}^2$ and the Ge shell is $23.5  \ \mathrm {nm}^2$, and lengths of both core and shell are $120 \ \mathrm {nm}$. In the Ge/Si case we use the same numbers of atoms for the core and shell, with the two materials exchanged. 
Note that due to the different lattice constant of Si and Ge, reaching exactly same area for two shapes is impossible but the total difference in whole volumetric size is near 2\%. This difference had been included in calculation of thermal conductivity.

\begin{figure}
    \centering
    \includegraphics[width=1\linewidth]{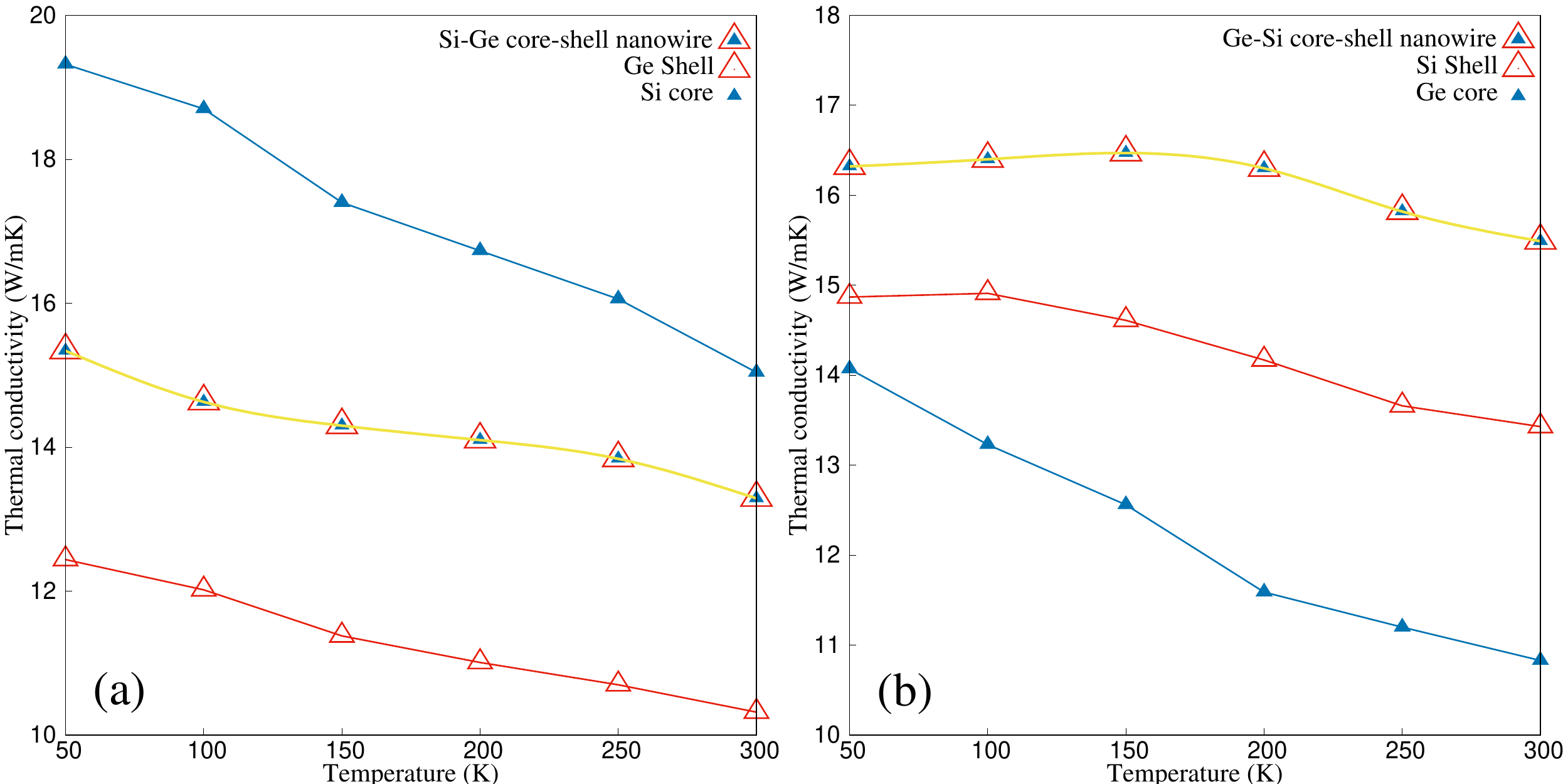}
    \caption{ Thermal conductivity variation with temperature for core/shell nanowires and separated solid nanowire and tubular nanowires. (a) Si/Ge core/shell nanowire and Si solid nanowire and tubular Ge nanowire. (b) Ge/Si core/shell nanowire and Ge solid nanowire and tubular Si nanowire.}
    \label{core/shell}
\end{figure}

In Fig.\ \ref{core/shell} (a) we show the thermal conductivity vs. temperature of the Si/Ge example. Despite having less atoms in the Si core than in the Ge shell, the isolated Si core shows higher thermal conductivity than the isolated Ge shell, as can also be seen in Table. 1.  This is because of two main reasons. First, the hole inside the tubular Ge nanowire lead to a large reduction of the thermal conductivity, because of the large surface to volume ratio and consequently intense phonons scattering. Second, in general Si nanowires have higher thermal conductivity than Ge nanowires with similar geometric parameters \citep{crismari2012thermal,mingo2003predicting}. 
{\blue The depression of the vibrational density of states of Si atoms on the surface with the addition of Ge shell atom leads to the reduction of the thermal conductivity of Si/Ge nanowire with respect to the Si core alone, as already shown by Hu et al.\cite{Hu2010}. However, now we observe that the difference between the average conductivity ($\kappa_{ave}$) and the true value for the core/shell structure become smaller with increasing the temperature, and reverse order for temperatures about 170 K. 
}

In Fig.\ref{core/shell} (b) we show the Ge/Si core/shell case, where  the thermal conductivity of this structure is larger than for the Ge and Si separated components. Despite the hole in the Si shell, which reduces the thermal conductivity if we compare it to the solid one, it still remains larger than the thermal conductivity of the Ge core. Because of this, Ge/Si core/shell nanowires shows higher thermal conductivity than Si/Ge in all temperature regime. This inequality has been obtained before for the square geometry \citep{crismari2012thermal,nika2013thermal}. This difference is not huge because despite carrying large amount of heat through both core and shell in Ge/Si core/shell nanowires, we still have shell atoms depressing over core atoms from vibrational states.


%
%





\section{Conclusions}

We have studied the thermal conductivity of solid and tubular silicon nanowires with different cross-sectional area, lengths and geometries. For silicon nanowires the thermal conductivity increases linearly by increasing the cross sectional area, and increases with a power law in the length of nanowires. By considering the same length and cross sectional area for all silicon nanowires with different cross sectional geometries, the square shape shows the lowest thermal conductivity value and the triangular shape shows the highest value.
Due to the increased surface to volume ratio, making an empty space inside nanowire, i.e. making a tubular nanowire, always leads to a decrease of the thermal conductivity. Tubular nanowires with thinner walls show less thermal conductivity and for the triangular shape the thermal conductivity decreases much more than for other geometries when the nanowires become thinner. 

{\blue We also calculated the thermal conductivity of core/shell Si/Ge and Ge/Si nanowires. In this case, due to phonon scattering at the interface between the two materials, the thermal conductivity of the nanowire is different from a simple combination of the core and shell thermal conductivity values calculated separately. 

By considering the same geometry parameters, but exchanging the materials order in core and shell (Ge/Si instead of Si/Ge), we can see a different thermal conductivity. 
Using nanowires of a specific geometry, and combining different materials, may be a promising way to engineer the heat transport at nanoscale.}



\ack
This work was supported by the Icelandic Research Fund, Grant 195943.

\appendix


\renewcommand{\thefigure}{A\arabic{figure}}
\setcounter{figure}{0}

\renewcommand{\theequation}{A\arabic{equation}}
\setcounter{equation}{0}

\section{More on methodology}
Comparison of popular force fields for predicting thermal properties of the bulk Si with that of \emph{ab-initio} method. Here we briefly describe the method and simulation detail for calculation of these properties.

\subsection{Simulation cell}
{\blue Fig.~\ref{fig:geo_wire} and \ref{fig:geo_tube} show the perspective view of some nanowires utilized in the present study those have the same cross-section area. The black box indicates simulation cell and vacuum around the nanowires in transverse direction.  
The illustrations were obtained by Ovito visualization tool. It is worth mentioning that that nanowires axes were considered to be parallel to to [111] orientation of diamond lattice. This gives (111) palnes in the nanowire cross-section. As can be seen in the figures, triangle, hexagonal and diamond cross-section adjust very well with the (111) plane i.e.\ they give atomically smooth and faceted surfaces around the nanowire. However, four-fold symmetry does not exist in the (111) plane and thus for the square cross-section only horizontal surfaces become smooth surface.

\begin{figure}[hbt!]
    \centering
    \includegraphics[width=0.85\linewidth]{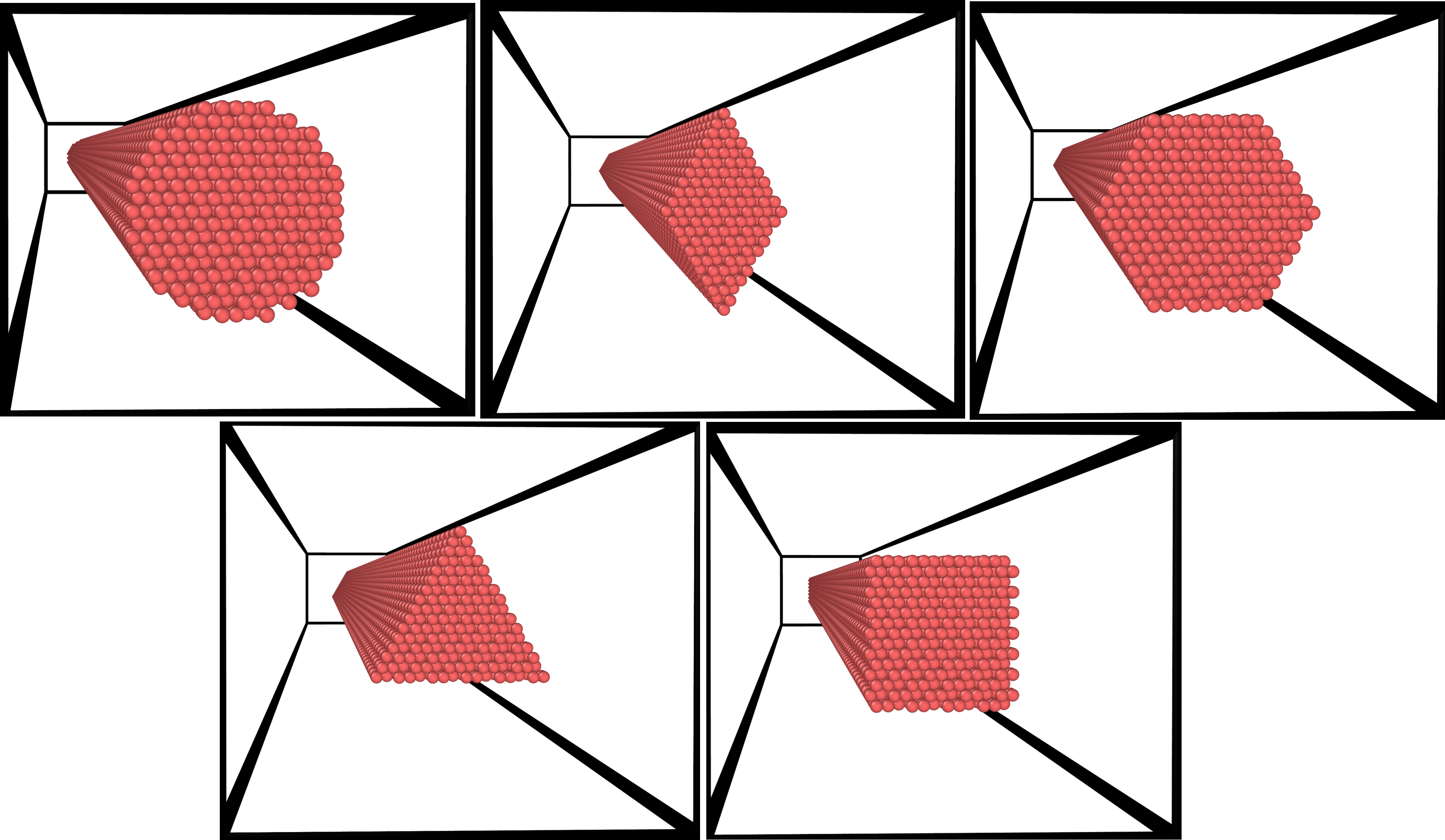}
    \caption{Cross sectional shapes of solid nanowires.}
    \label{fig:geo_wire}
\end{figure}
\begin{figure}[hbt!]
    \centering
    \includegraphics[width=0.85\linewidth]{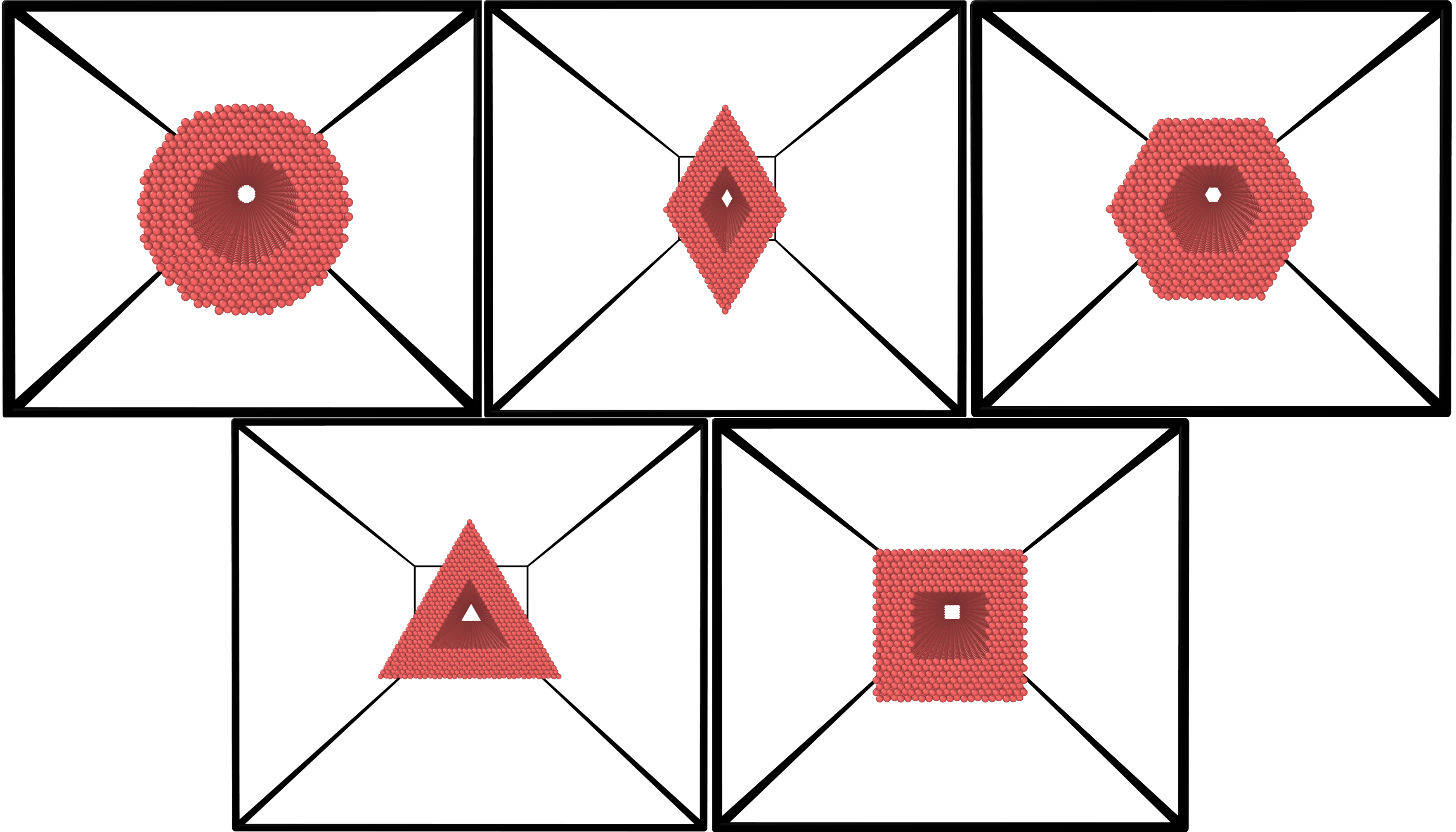}
    \caption{Cross sectional shapes of tubular nanowires.}
    \label{fig:geo_tube}
\end{figure} }

\subsection{First principles}
In the first principles approaches, plane wave projector augmented wave method \citep{blochl1994} was utilized within density functional theory framework as implemented in the VASP code \citep{kresse1999}. The generalized gradient approximation of Perdew Burke Ernzerhof \citep{pedrow1996} was utilized as exchange correlation potential. We used a plane-wave energy cutoff of 500~eV. The reciprocal space of diamond structure was sampled by 16$\times$16$\times$16 k-point mesh. The radial cutoffs of the PAW potentials of Si was 1.01~{\AA}. The 3s and 3p electrons of Si were treated as valence and the remaining electrons were kept frozen. The Brillouin zones of the unit cells were sampled by a 16x16x2 k-point mesh, and the Methfessel-Paxton scheme \citep{methfessel1989} with a smearing width of 0.4~eV was employed.

\subsection{Force field}

The general Tersoff \cite{tersoff1988new} potential takes the following form:
\begin{equation}
    U(r_{ij})=f_c(r_{ij})\big[A_{ij}\exp(-\lambda_{ij}r_{ij})-b_{ij}B_{ij}\exp(-\alpha_{ij}r_{ij})\big]
	\label{eq:short}
\end{equation}
with $A_{ij}$, $B_{ij}$, $\lambda_{ij}$ and $\alpha_{ij}$ being fitting parameters and $f_c$ being the smoothing function that works near the cuoffs. The main bond order term of Tersoff potential is $b_{ij}$ that changes the attraction based on the bond angle, number of nearest neighbors and their symmetry
\begin{equation}
    b_{ij}=\big[1+(\beta\zeta_{ij})^n\big]^{-\frac{1}{2n}}
\end{equation}
\begin{equation}
    \zeta_{ij}=\sum f_c(r_{ij})g(\theta_{ijk})\exp\big[\lambda^m(r_{ij}-r{ik})^m]
\end{equation}
\begin{equation}
    g(\theta_{ijk})=\Big(1+\big(\frac{c}{d}\big)^2-\frac{c^2}{d^2+(h-\cos\theta_{ijk})^2}\Big)\gamma_{ijk}
\end{equation}
here $\alpha$, $\beta$, $n$, $m$, $c$, $d$, and $h$ are fitting constants. It is worth mentioning that this is the general formalism of Tersoff potential and some of these parameters are known constants depending on the specific formalism. 

\subsection{Phonons}
Phonon calculations were performed by the supercell approach. Real-space force constants of supercells were calculated in the density-functional perturbation theory (DFPT) implemented in the VASP code \citep{kresse2013}, and phonon frequencies were calculated from the force constants using the PHONOPY code \citep{togo2008}. Thermodynamic properties were calculated using the quasi-harmonic approximation (QHA). For the QHA calculations, supercells containing 4$\times$4$\times$4 unit cells. It is worth mentioning that the difference between the DFPT the Parlinski-Li-Kawazoe method \citep{parlinski1997} with a finite displacement is negligible. However, the difference between supercell sizes increase with increasing temperature. For instance, the energy differences between 1$\times$1$\times$1 and 4$\times$4$\times$1 supercells, at 1000~K is several hundreds of meV/unit cell \citep{togo2010}.

\subsection{Thermodynamic properties}
According to QHA the thermal properties of solids at constant volume can
be calculated from their phonon density of states as a function of frequencies \citep{togo2010}. The phonon contribution to the Helmholtz free energy $F_{\rm phonon}$ is given by
\begin{equation}
    F_{\rm phonon}=\frac{1}{2}\sum_{\mathbf{q},v}\hbar\omega_{\mathbf{q},v}+k_{\rm B}T\sum_{\mathbf{q},v}\ln\Big[1-\exp\Big(\frac{\hbar\omega_{\mathbf{q},v}}{k_{\rm B}T}\Big)\Big]
\end{equation}
where $\mathbf{q}$ and $v$ are the wave vector and band index, respectively, $\omega_{\mathbf{q},v}$ is the phonon frequency at $\mathbf{q}$ and $v$, and $T$ is the temperature. $k_{\rm B}$ and $\hbar$ are the Boltzmann constant and the reduced Planck constant, respectively.

The heat capacity ($C_V$) and the entropy $S$ at constant volume are given by
\begin{equation}
    C_V=\sum_{\mathbf{q},v}k_{\rm B}\Big(\frac{\hbar\omega_{\mathbf{q},v}}{k_{\rm B}T}\Big)^2\frac{\exp(\hbar\omega_{\mathbf{q},v}/k_{\rm B}T)}{[\exp(\hbar\omega_{\mathbf{q},v}/k_{\rm B}T)-1]^2}
\end{equation}
and
\begin{equation}
    S=-k_{\rm B}\sum_{\mathbf{q},v}\ln\Big[1-\exp\Big(\frac{\hbar\omega_{\mathbf{q},v}}{k_{\rm B}T}\Big)\Big]-\frac{1}{T}\sum_{\mathbf{q},v}\frac{\hbar\omega_{\mathbf{q},v}}{\exp(\hbar\omega_{\mathbf{q},v}/k_{\rm B}T)-1}
\end{equation}
respectively.

In practical thermodynamical problems related to solids, the thermal properties need to be known at constant pressure. They can be calculated from the previous quantities through thermodynamic relationship. The Gibbs free energy ($G$) may be written as
\begin{equation}
    G(T,P)=\min_{V} [U(V)+F_{\rm phonon}(T;V)+PV],
    \label{eq:gibbs}
\end{equation}
where $V$ and $P$ are the volume and pressure, respectively, and $U(V)$ is the total energy of electronic structure at constant volume. The right-hand side of Eq.~(\ref{eq:gibbs}) means that, for each couple of $T$ and $P$ variables, the function inside the square brackets is minimized with respect to the volume. Then the heat capacity at constant pressure is derived from $G(T,P)$ by
\begin{equation}
\hspace{-20mm}    C_P(T,P)=-T\frac{\partial^2G(T,P)}{\partial T^2}=T\frac{\partial V(T,P)}{\partial T}\frac{\partial S(T;V)}{\partial V}|_{V=V(T,P)}+C_V[T,V(T,P)]
\end{equation}
where $V(T,P)$ is the equilibrium volume at $T$ and $P$.

\subsection{Comparison of different interatomic potentials}

\begin{figure}[H]
    \centering
    \includegraphics[width=1\linewidth]{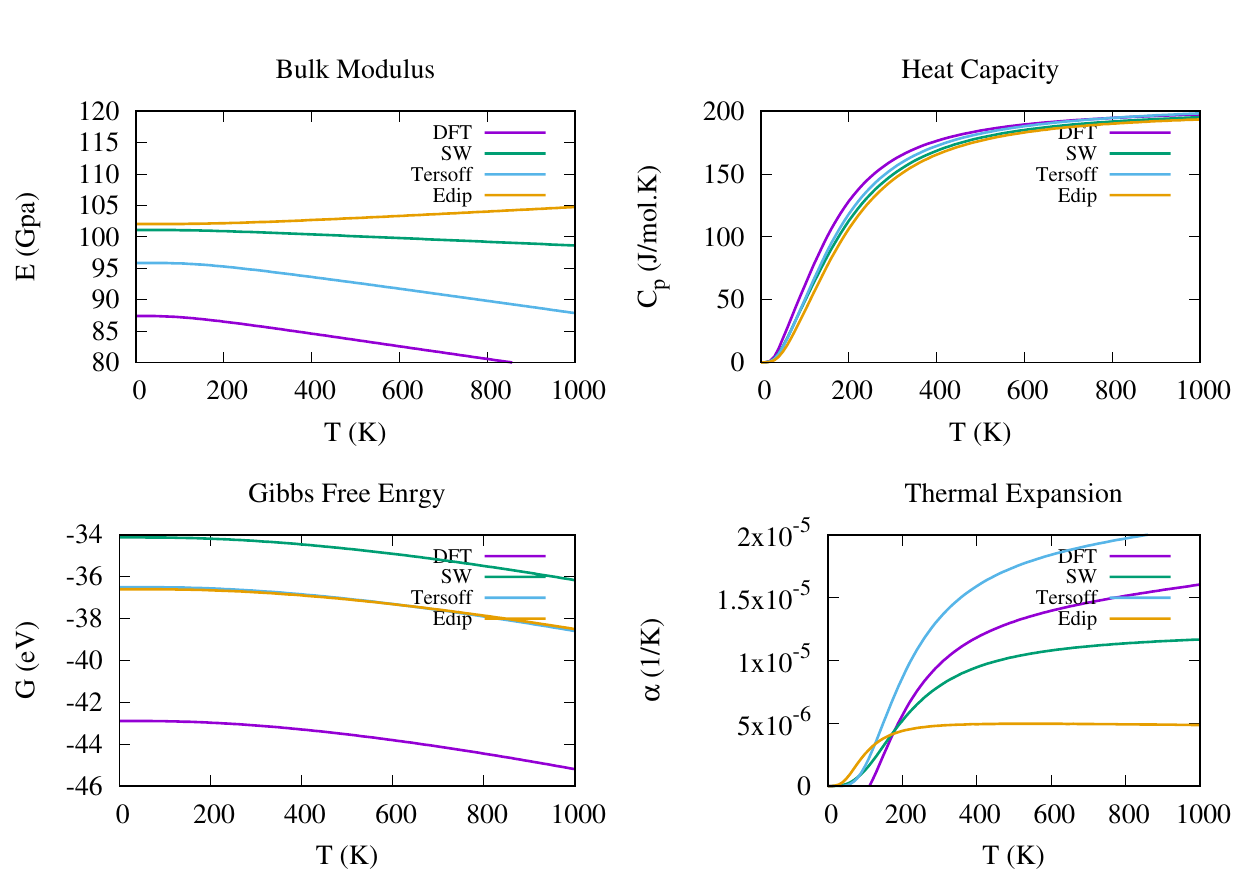}
    \caption{Variation of some important properties with temperature for Tersoff, SW and EDIP potential compared to that of DFT.}
    \label{fig:pots}
\end{figure}

Fig.~\ref{fig:pots} shows the comparison of various thermodynamic properties obtained by two popular (SW and Tersoff) potentials with a more recent EDIP potential and DFT results. The bulk modulus ($E$) is calculated by fitting attractive regions of energy vs.\ volume (E-V) curves (E-V curves are not shown here). Rest of the plots were calculated from phonon dispersion obtained by supercell approach within second harmonic approximation theorem. It can be seen that Gibbs free energy and $E$ and heat capacity obtained by Tersoff potential shows better agreement with DFT results compared to that of SW. The only exception is thermal expansion coefficient  ($\alpha$) in which SW potential predicts closer results to that of DFT. At elevated temperatures EDIP potential predicts increasing trend of the bulk elastic modulus and consequently underestimates $\alpha$. At temperatures below 130~K, DFT results indicates negative $\alpha$ that is not predicted by interatomic potential. 
%


\bibliographystyle{iopart-num}
\bibliography{Bibliography2}

\end{document}